\documentclass[aps,prl,twocolumn,superscriptaddress,amsfont,graphicx,nofootinbib,preprintnumbers]{revtex4-1}%
\usepackage{color,graphicx,epsfig}
\usepackage{ifpdf}
\usepackage{amsmath}
\usepackage{bm}
\usepackage{color}
\usepackage[english]{babel}
\usepackage{graphicx}%
\usepackage{amsfonts}%
\usepackage{amssymb}
\usepackage{braket}
\usepackage{hyperref}
\usepackage{enumerate}
\usepackage{comment}

\definecolor{nicered}{rgb}{0.7,0.1,0.1}
\definecolor{nicegreen}{rgb}{0.1,0.5,0.1}
\hypersetup{colorlinks,citecolor= nicegreen,linkcolor= nicered}

\usepackage{graphicx}
\usepackage{dcolumn}
\usepackage{bm}
\usepackage{slashed}

\begin{document}

\title{Light gravitino dark matter: LHC searches and Hubble tension}

\author{Yuchao Gu}
\affiliation{Department of Physics and Institute of Theoretical Physics, Nanjing Normal University, Nanjing, 210023, China}
\affiliation{School of Physics, Yantai University, Yantai 264005, China}

\author{Maxim Khlopov}
\affiliation{Université de Paris, CNRS, Astroparticule et Cosmologie, F-75013 Paris, France}
\affiliation{Centre for Cosmoparticle Physics Cosmion; National Research Nuclear University MEPHI (Moscow Engineering Physics Institute), Kashirskoe Sh., 31, Moscow 115409, Russia}
\affiliation{Institute of Physics, Southern Federal University, Stachki 194, Rostov on Don 344090, Russia}

\author{Lei Wu}
\affiliation{Department of Physics and Institute of Theoretical Physics, Nanjing Normal University, Nanjing, 210023, China}

\author{Jin Min Yang}
\affiliation{Department of Physics, Tohoku University, Sendai 980-8578, Japan}
\affiliation{CAS Key Laboratory of Theoretical Physics, Institute of Theoretical Physics, Chinese Academy of Sciences, Beijing 100190, China}
\affiliation{School of Physics, University of Chinese Academy of Sciences, Beijing 100049, China}

\author{Bin Zhu}
\affiliation{School of Physics, Yantai University, Yantai 264005, China}
\affiliation{Department of Physics, Chung-Ang University, Seoul 06974, Korea}

\date{\today}

\begin{abstract}
The recent measurements of the cosmological parameter $H_0$ from the direct local observations 
and the inferred value from the Cosmic Microwave Background show $\sim 4 \sigma$ discrepancy. This may indicate new physics beyond the standard $\Lambda$CDM.
We investigate the keV gravitino dark matter that has a small fraction of non-thermal component
(e.g. from the late decay of NLSP bino) under various cosmological constraints. We find such a scenario is highly predictive and can be tested by searching for the dilepton plus missing energy events at the LHC. Besides, we also discuss its implication for Hubble tension, however, which can be reduced to $3\sigma$ level marginally. 
\end{abstract}
\pacs{Valid PACS appear here}
\maketitle


\section{Introduction}
The $\Lambda$CDM model combining cold dark matter (CDM) with a cosmological constant $\Lambda$ is remarkably 
successful in describing the results of cosmological observations. However, recently there is a growing 
tension in the determinations of Hubble constant, for example, the measurement from Cepheid-calibrated Type 
Ia Supernova $H_0=(74.03 \pm 1.42)$ km s$^{-1}$ Mpc$^{-1}$~\cite{Riess:2019cxk} shows about 
$4.4\sigma$ discrepancy with the inferred value 
$H_0=(67.36 \pm 0.54)$ km s$^{-1}$ Mpc$^{-1}$~\cite{Aghanim:2018eyx} from the Cosmic Microwave Background (CMB).
Due to the size of the discrepancy and the independence of the observations, a single systematic error 
in the data seems impossible to completely solve the discrepancy~\cite{Verde:2019ivm}. Therefore, the $H_0$ 
tension may call for new physics beyond the standard $\Lambda$CDM~(see e.g.~\cite{Berezhiani:2015yta,Bernal:2016gxb,Zhao:2017urm,Raveri:2017jto,Buen-Abad:2018mas,DEramo:2018vss,Guo:2018ans,Poulin:2018cxd,Vattis:2019efj,Choi:2019jck,Alcaniz:2019kah,Feng:2019jqa,Kreisch:2019yzn,Agrawal:2019lmo,He:2020zns,Lyu:2020lps,Choi:2020tqp} and reference therein).

As known, the cold dark matter can be an explanation for the formation of large-scale structure and galaxies. 
Despite of its success, the predictions made by the CDM deviate from the observational data in small-scale 
structure, such as \emph{core-cusp}~\cite{Walker:2011zu}, \emph{missing satellite}~\cite{Klypin:1999uc} 
and \emph{too big to fail}~\cite{BoylanKolchin:2011de} problems. One possible way of solving these problems 
is to introduce warm dark matter particles.
The free-streaming motion of such warm DM particles 
reduce power on small scales, but keep the CDM predictions for the formation of large-scale 
structure. Besides, many models of dark matter in particle physics are not always consisting of the pure CDM. 
Thus, in conjunction with the Hubble tension, it seems timely to explore the possibilities of departures 
from the standard CDM model.

Also note that, as a compelling dark matter candidate, 
the Weakly Interacting Massive Particle (WIMP)~\cite{Lee:1977ua} 
has been searched for in various (in)direct detections~\cite{Jungman:1995df} and 
collider experiments~\cite{Buchmueller:2017qhf}. However, the null results of detection have produced 
stringent bounds on such interactions, which have motivated to explore the dark matter at lower masses 
and/or with different detection signatures~(for recent reviews, see e.g.~\cite{Knapen:2017xzo,Bertone:2018krk}).

If DM particle is sufficiently light, it may affect the radiation energy density by mimicking an additional neutrino species in the early 
universe~\cite{Hooper:2011aj}.  During the radiation era, the neutrino energy density $\rho_\nu$ in flat geometry is related with the Hubble constant $H(t)$ by
\begin{eqnarray}
H^2(t) \simeq \frac{8\pi G}{3}(\rho_\gamma+\rho_\nu).
\end{eqnarray}
where $\rho_\gamma$ is the photon energy density. Any process that changes the abundance of neutrinos can alter the expansion rate of the universe. Interestingly, the Hubble constant and the effective number of neutrino species can have a positive correlation because the non-standard $N_{\rm eff}$ can affect the sound horizon, which in turn changes the angular position of the acoustic peaks.

In this paper, we study the light gravitino ($\tilde{G}$) dark matter that is always predicted by locally supersymmetric extensions of the SM~\cite{Wess:1992cp}. Depending on the 
supersymmetry-breaking mechanisms, the gravitino mass can range from eV scale up to the scale beyond 
the TeV region~\cite{Pagels:1981ke,Weinberg:1982zq,Khlopov:1984pf,Dine:1994vc,Giudice:1998bp,Randall:1998uk,Ellis:2003dn,Buchmuller:2005rt,Dudas:2017rpa}. 
If the gravitino is the lightest supersymmetric 
particle (LSP), it can play the role of dark matter particle, which may or may not be in the thermal equilibrium 
with the hot primordial plasma. When the gravitino dark matter is light enough and non-thermally produced from 
the late decay of the heavier next-to-LSP (NLSP), it can contribute to the radiation density by mimicking an 
extra neutrino species. Therefore, such a light gravitino dark matter may reduce the Hubble tension. More interestingly, 
our framework is highly predictive and can be tested by the LHC experiment.

\section{Light Gravitino Dark Matter}
 
The gravitino is present in the gauge theory of local supersymmetry. It is the spin-$3/2$ superpartner 
of the graviton. The gravitino interactions are determined by supergravity and by the MSSM parameters 
and are suppressed by the Planck mass. The gravitino mass is obtained via the Super-Higgs mechanism~\cite{Cremmer:1982en}  and strongly depends on the SUSY breaking schemes. In the gauge mediated supersymmetry-breaking (GMSB) models, 
the gravitino is usually the LSP and has a mass in the range of 1 eV $ \lesssim m_{3/2} \lesssim$ 1 GeV~\cite{Giudice:1998xp}. However, this light gravitino dark matter may lead to some cosmological problems~\cite{Moroi:1993mb,Asaka:2000zh,Bolz:2000fu,Roszkowski:2004jd,Cerdeno:2005eu,Pradler:2006qh,Pradler:2006hh}. For example, if the gravitino was thermalized in the early universe, 
its mass $m_{3/2}$ should be less than  $\sim 1$ KeV to avoid overclosing the universe. Otherwise, a low reheating 
temperature of inflation $T_R$ is required to dilute the gravitino abundance and thus fails to explain the baryon 
asymmetry by the thermal leptogenesis.   

On the other hand, the messenger particles are always predicted by the GMSB models, whose superpotential 
is usually given by
\begin{equation}
W=S\Phi_{M}\bar{\Phi}_M + \Delta W(S,Z_i),
\end{equation}
where $S$ and $Z_i$ are respectively the spurion left chiral superfield and the secluded sector fields, 
and $\Phi_M$ and $\bar{\Phi}_M$ are the messenger left chiral superfields which are charged under the SM gauge 
group and transmit the SUSY breaking effect to the visible sector in terms of gauge interaction at the loop level. 
In the minimal version of the GMSB, the messenger number is conserved so that the lightest messenger particle 
would easily overclose the universe, unless it can be diluted to a very low abundance or has a tens of TeV mass. 
However, it should be noted that the lightest messenger can have interactions with the SM particles and sparticles 
by introducing additional messenger-matter interactions or gauge interaction~\cite{Baltz:2001rq,Fujii:2002fv,Jedamzik:2005ir}. Then, the late decay of the lightest messenger to visible sector particles can produce a substantial 
amount of entropy, and will dilute the light gravitino relic density to the observed value in the present universe. 
The dilution factor arising from the messenger decay can be parameterized by
\begin{equation}
D_m=\frac{4/3 M_m Y_m}{(90/g_*\pi^2)^{1/4}\sqrt{\Gamma_m M_P}},
\end{equation}
where $Y_m$ is the yield of lightest messenger, $M_m$ is the mass of messenger and $\Gamma_m$ is the 
messenger decay width, and $g_{*}$ denotes the number of relativisitic degree of freedom at the temperature 
of the lightest messenger decay. 

Given that the gravitino couplings are extremely weak, the pre-existing gravitino can be in or out of the 
thermal equilibrium in the early universe. The  freeze-out temperature of the gravitino $T^{3/2}_f$ is given by 
\begin{eqnarray}
 T^{3/2}_{f} && \approx 0.66{\rm TeV}~\left(\frac{g_{*}}{100}\right)^{1 / 2}\left(\frac{m_{3 / 2}}{10 {\rm keV}}\right)^{2}\left(\frac{1 {\rm TeV}}{m_{\tilde{g}}}\right)^{2},
 \label{fz}
\end{eqnarray}
where $g^*$ is the effective degrees of freedom of relativistic particles at the gravitino freeze-out temperature and has the value in the range of 90-140~\cite{Pierpaoli:1997im}. $m_{\tilde{g}}$ is the mass of gluino and should be heavier than 1 TeV according to the current LHC limits. From Eq.~\ref{fz}, it can be seen that a keV gravitino corresponds to a low freeze-out temperature 
$T^{3/2}_f \sim 10$ GeV. On the other hand, thanks to the messenger dilution effect, the reheating temperature 
$T_R$ can be as high as $\sim 10^{9}$ GeV for the thermal leptogenesis. This indicates that such a light gravitino 
dark matter in the GMSB should be thermalized in the early universe, and its relic density can be calculated by
\begin{equation}
\Omega_{3 / 2}^{\rm TP} h^{2}=1.14\left(\frac{g_{*}}{100}\right)^{-1}\left(\frac{m_{3 / 2}}{\mathrm{keV}}\right).
\label{eqn:hot}
\end{equation}
Note that the gravitino can also be non-thermally generated via the late decay of the NLSP, for examlpe the 
radiative decay of bino, $\tilde{B} \to \tilde{G} \gamma$~\cite{Feng:2003xh,Feng:2003uy,Feng:2004mt}. As stated above, 
such a non-thermal gravitino dark matter may be a solution to the Hubble constant problem. The non-thermal relic 
density of gravitino is given by
\begin{eqnarray}
\Omega_{3 / 2}^{{\rm NTP}} h^{2}&=&m_{3 / 2} Y_{\tilde B}\left(T_{0}\right) s\left(T_{0}\right) h^{2} / \rho_{c}\nonumber \\
&=&\frac{m_{3 / 2}}{m_{\tilde B}} \Omega_{\tilde B} h^{2},
\end{eqnarray}
with
\begin{eqnarray}
\Omega_{\tilde{B}}h^2&&=0.0013\left(\frac{m_{\tilde{\ell}_{R}}}{100 {\rm GeV}}\right)^2 \frac{\left(1+R\right)^4}{R(1+R^2)} \nonumber \\
&& \times \left(1+0.07\log\frac{\sqrt{R} 
\times 100 {\rm GeV}}{m_{\tilde{\ell}_{R}}}\right)
\label{nlsp}
\end{eqnarray}
where the mass ratio $R \equiv m^2_{\tilde{B}}/m^2_{{\tilde{\ell}_{R}}}$. Given the strong LHC bounds on the 
squarks and gluinos, we only include the contributions of the right-handed sleptons to the relic abundance 
of bino NLSP in Eq.~\ref{nlsp}~\cite{ArkaniHamed:2006mb}. For simplicity, we assume $m_{\tilde{\ell}_{R}}$ as 
a common mass parameter of the three generation right-handed sleptons. It should be mentioned that only the 
first-two generation sleptons should be included in Eq.~\ref{nlsp} when $m_{\tilde{B}}$ is less than $m_\tau$.

Since the decay width of the lightest messenger is much smaller than the gravitino freeze-out temperature $T_f^{3/2}$, 
the messenger decay can dilute the thermally produced gravitinos. Besides, the freeze-out temperature of the bino 
NLSP is usually $\sim m_{\tilde B}/20$. If the bino mass is around 1 GeV, it can still freeze out before the messenger 
decay and then be diluted by the entropy production. It should be noted that the non-thermally produced gravitinos
from the bino late decay  
will not be further diluted as long as the bino decay is sufficiently delayed. Therefore, the final gravitino 
abundance can be calculated by
\begin{equation}
\Omega_{3/2}h^2=\frac{1}{D_m}(\Omega_{3/2}^{\mathrm{TP}}h^2+\Omega_{3/2}^{\mathrm{NTP}}h^2).
\end{equation}
In our study, we require that the gravitinos solely compose the dark matter and satisfy the observed relic density 
within the $3\sigma$ range, $0.075<\Omega_{3/2}h^2<0.126$~\cite{Ade:2015xua}.  

Another benefit of the messenger decay in our scenario is that the entropy production can cool down the velocity 
of the thermally produced gravitino dark matter. For example, when a particle with mass $m$ freezes out from 
the primordial plasma relativistically, it has a present-day velocitiy 
$\langle v^{0}_{3/2} \rangle \approx 0.023 \mathrm{kms}^{-1}\left(g_*(T_{\rm dec}) / 100\right)^{-1 / 3}(m / 1 \mathrm{keV})^{-1}$, which will be reduced to $\sim \langle v^{0}_{3/2} \rangle /D^{1/3}_m $. 
Depending on the dilution factor, the thermally produced gravitino may become non-relativistic, even its mass 
is less than $\sim$ 10 keV. Whereas, the non-thermally produced gravitino that inherits the kinetic energy 
from the bino decay can be still relativistic. Due to the vague limits between hot, warm and cold dark matter, 
we identify the thermal gravitino dark matter as the CDM when 
$\langle v_{3/2} \rangle < 0.1\langle v^0_{3/2} \rangle$ in the following calculations. 

\section{Constraints}

The gravitino dark matter from the late decay of the bino can be nearly relativistic, and thus produce an extra 
radiation density $\rho^{\rm extra}_R=f \times \rho_{3/2} \times (\gamma_{3/2}-1) $ in the early universe, 
where $f=\Omega_{3/2}^{\mathrm{NTP}} h^2/ (\Omega_{3/2}^{\mathrm{TP}} h^2+\Omega_{3/2}^{\mathrm{NTP}} h^2) $ is the fraction of 
the non-thermal gravitino density in the total gravitino production and $\gamma_{3/2}$ is the boost factor of the 
gravitino from the bino decay. At the matter-radiation equality, the energy density per neutrino species is approximately equal 
to 16\% of the energy density of CDM. This implies that the non-thermal gravitino dark matter that has a kinetic energy equivalent 
to 1.16 can be regarded as an additional neutrino species. Therefore, the resulting effective neutrino 
species $\Delta N_{\mathrm{eff}}$ can be given by~\cite{Hooper:2011aj}
\begin{equation}
\Delta N_{\mathrm{eff}}=f \times\left(\gamma_{3/2}-1\right) / 0.16
\end{equation}
with
\begin{eqnarray}
\gamma_{3/2}(a)&=&1+\left(\frac{a_{\tau}}{a}\right)\left(\frac{m_{\tilde B}}{2 m_{3/2}}+\frac{m_{3/2}}{2 m_{\tilde B}}-1\right)
\end{eqnarray}
where $a_\tau$ is the scale factor at the time of bino decay. In Ref.~\cite{Bernal:2016gxb}, a comprehensive investigation of the CMB data and direct measurements shows a positive correlation between $N_{\rm eff}$ and $H_0$. For example, when $0.29<\Delta N_{\rm eff}<0.85$, the Hubble constant can reach $H_0=74.03$ km s$^{-1}$ Mpc$^{-1}$. Similar results are also given in~\cite{Vagnozzi:2019ezj}. Thus, increasing the effective number of neutrino species may provide an avenue to ameliorate the Hubble tension. However, it should be noted that the extra relativistic degree of freedom is constrained by the Planck data and baryon acoustic oscillation (BAO) data, which indicates $N_{\rm eff} = 2.99 \pm 0.17$~\cite{Aghanim:2018eyx}. Comparing with the prediction 
$N_{\rm eff}=3.046$~\cite{Mangano:2005cc,Akita:2020szl} from the Standard Model (SM) with three generations of fermions, this produces an upper bound $\Delta N_{\rm eff} < 0.29$ at 95 \% C.L.. We will include this constraint in our following numerical calculations.

On the other hand, we should consider that the non-thermal gravitino will affect the growth of the structure 
due to its large free-streaming length. The free-streaming starts at the bino decay time and 
finishes at matter-radiation equality, which is given by
\begin{eqnarray}
\lambda_{\mathrm{FS}}&=&\int_{\tau}^{t_{\mathrm{eq}}} \frac{v_{3/2}(t)}{a(t)} d t \nonumber \\
&\simeq& 0.6 \mathrm{Mpc}\times \left(\frac{m_{\tilde B}}{10 m_{3/2}}\right)\left(\frac{\tau}{10^{4} \mathrm{sec}}\right)^{1 / 2}\nonumber \\
&\times& \left[1+0.1 \log \left(\frac{10 m_{3/2}}{m_{\tilde B}}\left(\frac{10^{4} \mathrm{sec}}{\tau}\right)^{1 / 2}\right)\right].
\end{eqnarray}
If the free-streaming distance that the gravitino propagates is larger than $\sim$ Mpc set by the Lyman-alpha forest~\cite{Viel:2010bn}, it roughly cannot form the observed large-scale structure which in turn puts 
a constraint on the non-themral gravitino dark matter. By fitting the CMB data~\cite{Aghanim:2018eyx,Fidler:2018bkg}, 
the large-scale structure observations~\cite{Viel:2010bn} and cosmological simulations~\cite{Liu:2017now}, 
it is found that the fraction of the non-thermal gravitino dark matter has to be very small. 
In order to suppress such a contribution, one can require the distortion on the linear matter power 
spectra ${\rm exp}(-4.9f)>0.95$~\cite{Ma:1996za,Eisenstein:1997jh}, which corresponds to $f<0.01$. 

Besides, the late decay of bino via the process $\tilde{B} \to \tilde{G}\gamma$ may affect the big-bang 
nucleosynthesis (BBN)~\cite{Yang:1983gn}, whose life-time in the limit of $m_{\tilde{B}} \gg  m_{\tilde{G}}$ is approximately given by
\begin{eqnarray}
\tau_{\tilde{B}} \simeq \frac{48 \pi M_{P}^{2}}{\cos^2\theta_W}\left(\frac{m_{3/2}^2}{m_{\tilde{B}}^5}\right) .
\end{eqnarray} 
The photons from the bino decay may induce electromagnetic showers through their scattering off the background 
photons and electrons~\cite{Cyburt:2009pg,Kawasaki:2017bqm}. The energetic photon in the shower can destroy 
the light elements such as D and $^4$He. The photodissociation of $^4$He happens at the cosmic time 
of $\gtrsim 10^{6}$s, while photodissociation of D will be important at higher temperature because of the 
smallness of its binding energy, which corresponds to a long-lived particle with a lifetime longer 
than $10^{4}$s~\cite{Holtmann:1998gd,Kawasaki:2000qr}. Thus, we require the life-time of our late decaying 
bino is shorter than $10^{4}$s to avoid the BBN constraints. 

\section{Numerical Results and Discussions}
Next, we perform a numerical study to explore the allowed parameter space of our scenario. There are only four relevant input parameters in our scenario: $m_{\tilde{B}}$, $m_{\tilde{G}}$, $m_{\tilde{\ell}_R}$ and $D_m$. We consider all the constraints stated above. 

\begin{figure}[ht]
\centering
\includegraphics[width=8.5cm,height=7.5cm]{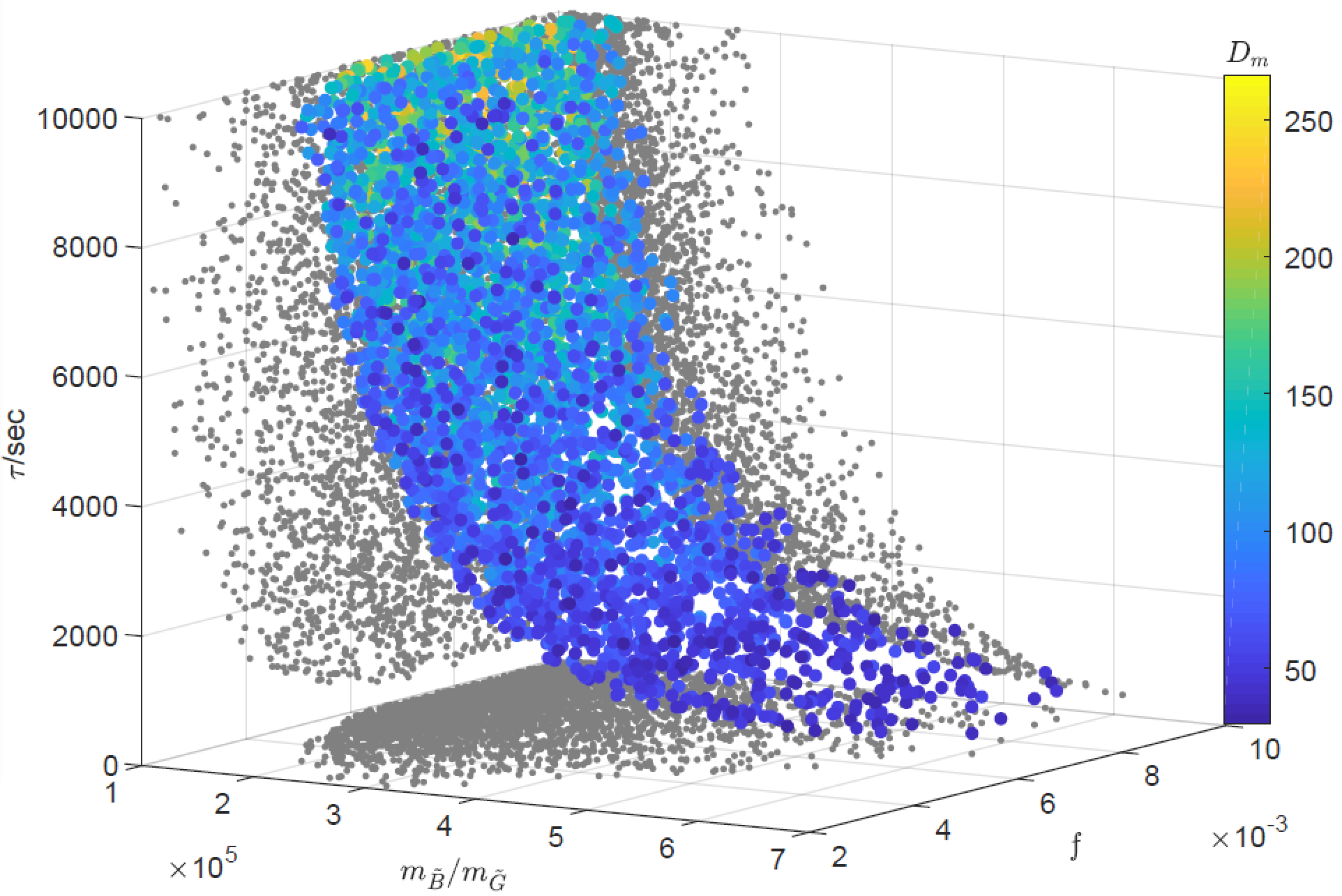}
\caption{Samples satisfying the DM relic density ($0.075<\Omega h^2<0.126$), the BBN constraint ($\tau_{\tilde{B}}<10^4$s) and other cosmological observations ($f<0.01$ and $N_{\rm eff}<0.29$). The colormap denotes the values of dilution factor.}
\label{fig:fit}
\end{figure}
In Fig.~\ref{fig:fit}, we present the results of the life-time of bino ($\tau$), the mass ratio 
$m_{\tilde{B}}/m_{\tilde{G}}$, the non-thermal gravitino DM fraction $f$ and the dilution factor $D_m$ 
for the samples allowed by the experimental constraints. It can be seen that there is a strong correlation 
between these quantities. The life-time of bino deceases as the mass ratio $m_{\tilde{B}}/m_{\tilde{G}}$ becomes large, 
which can be much smaller than the BBN bound. The dilution factor is required to be in the range 
of $29<D_m <266$. The non-thermal gravitino DM fraction $f$ can be suppressed to ${\cal O}(10^{-3})$. 
When $m_{\tilde{B}}$ is fixed, a light gravitino will lead to a small thermal relic density of the gravitino DM, 
while a heavy gravitino will need a large dilution factor to reduce the thermal relic density. 
Both cases can result in a large value of $f$.  On the other hand, for a given slepton mass, 
a heavy $m_{\tilde{B}}$ will increase the relic density of non-thermal gravitino DM and thus enhance 
the value of $f$.

We comment on the possible realization of a large dilution factor $D_m$ in Fig.~\ref{fig:fit}. 
For example, in the general gauge mediation, the messenger sector can be $5\oplus\bar 5$ representation 
under $SU(5)$. The interaction between messenger and matter fields in minimal Kahler potential can be written as~\cite{Fujii:2002fv}, 
\begin{equation}
    \delta K=\lambda \Phi_m \bar{5} 
\end{equation}
where $\lambda$ is an $\mathcal{O}(1)$ parameter from the naturalness criteria, and  $\bar{5}$ stands for matter fields in the SM. In terms of Kahler transformation, this interaction can be reinterpreted as additional interaction in superpotential,
\begin{equation}
    \delta W=\lambda m_{3/2} \Phi_m \bar{5}
\end{equation}
Assuming that only gauge interaction plays a crucial role, the decay width of messenger is approximately given by~\cite{Staub:2009ww},
\begin{equation}
\Gamma_m=\frac{g^2}{16\pi}\left(\frac{m_{3/2}}{\sqrt{2}M_m}\right)^2 M_m.
\end{equation}
Here it can be seen that the appearance of the small factor $(m_{3/2}/M_m)^2$ will naturally lead to a tiny decay width of the lightest messenger. When $M_m$ is about ${\cal O}(10^{8})$ GeV, the corresponding yield $Y_m$ is around $ {\cal O}(10^{-9})$ and the decay width of the lightest messenger is about ${\cal O}(10^{-22})$ GeV. 
Then, one can have a dilution factor of ${\cal O}(10)$-${\cal O}(10^2)$. As an example, we present a benchmark point in Table~\ref{tab:benchmarks}.


\begin{table}[htbp]
\begin{centering}
\begin{tabular}{|c|c|c|c|c|c|} \hline  
$M_m$ (GeV)        &  
$m_{32}$ (keV)    & 
$Y_m$         & 
$\Gamma_m$ (GeV)  & 
$D_m$   \\ \hline   
$3\times 10^8$     &
5  &
$7.1\times 10^{-9}$  & 
$3.5\times 10^{-22}$     & 
100 \\ \hline
\end{tabular}
\caption{A benchmark point for the dilution factor $D_m$. }
\label{tab:benchmarks} 
\end{centering}
\end{table}

\begin{figure}[ht]
\centering
\includegraphics[width=8.5cm,height=7.5cm]{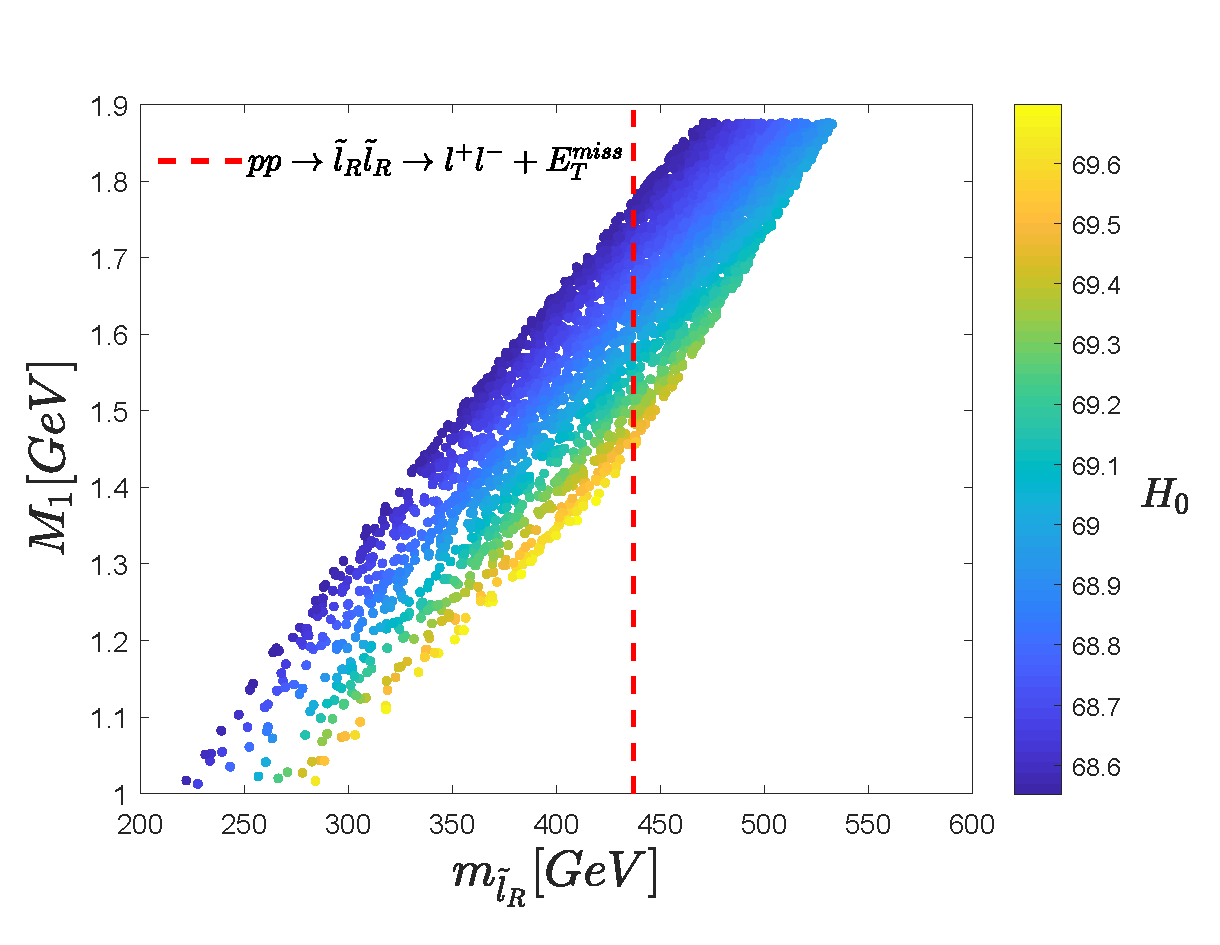}
\caption{Same as Fig.~\ref{fig:fit}, but projected on the plane of $m_{\tilde{B}}$ versus $m_{\tilde{\ell}_R}$. The colormap denotes the values of Hubble constant.
The dashed line correspond to the $95\%$ C.L. exclusion limits from the search for the slepton pair events 
at the 13 TeV LHC with a luminosity of ${\cal L}=139$ fb$^{-1}$~\cite{Aad:2019vnb}.}
\label{fig:lhc}
\end{figure}

In Fig.~\ref{fig:lhc}, we show the allowed samples on the plane of $m_{\tilde{B}}$ versus $m_{\tilde{\ell}_R}$. Following Ref.~\cite{Vagnozzi:2019ezj}, we calculate the corresponding values of Hubble constant for these samples and present their results as well. It can be seen that the Hubble constant can be mostly enhanced to about $H_0=69.69$, which reduces the tension to around $3\sigma$ level. It should be noted that the required effective neutrino number $\Delta N_{\rm eff}$ and the large-scale observation produces a lower and upper limit on the fraction $f$, respectively. 
These lead to the bounds on the slepton mass, since the non-thermal gravitino DM relic density depends 
on the abundance of bino. From Fig.~\ref{fig:lhc}, we can see that the slepton mass $m_{\tilde{\ell}}$ has to 
be less than about 520 GeV. Such a light slepton can be produced in pair through the Drell-Yan process 
$pp \to \tilde{\ell}^+\tilde{\ell}^-$ at the LHC. Due to the small mass splitting between the  bino and 
gravitino, the photon from the bino decay will be too soft to be observed by the detectors. 
Therefore, such slepton pair production process will give the dilepton plus missing energy signature at the LHC. 
In Fig.~\ref{fig:lhc}, we present the current LHC bounds of searching for selectron/smuon pair production, 
and find that the slepton with the mass lighter than about 440 GeV has been excluded. 
We can expect that the rest of parameter space can be fully probed by the HL-LHC.

\section{Conclusions}

In this paper, we studied the keV gravitino dark matter with a small fraction of non-thermal 
relic density in the gauge mediation supersymmetry breaking. Thanks to the messenger decay, the gravitino abundance 
can be diluted to the observed value, and also make the thermally produced gravitino still cold to satisfy 
the large-scale structure observations. We found that such a scenario can be tested by searching 
for slepton pair production at the LHC. Besides, since the non-thermal gravitino from the bino decay 
can mimic additional relativistic species, the expansion rate of the universe could be altered in the early universe. However, due to the current strong constraint on the effective neutrino number $\Delta N_{\rm eff}$, the Hubble tension can only be reduced to about $3\sigma$ level at best.

\section{acknowledgments}
We appreciate Feng Luo for his collaboration on early stage and thank Xin Zhang for his helpful discussions. This work is supported by the National Natural Science Foundation of China (NNSFC) under grant 
Nos. 117050934, 11847208, 11875179, 11805161,  11675242, 11821505, 11851303, 
by Jiangsu Specially Appointed Professor Program,
by Peng-Huan-Wu Theoretical Physics Innovation Center (11847612), 
by the CAS Center for Excellence in Particle Physics (CCEPP), 
by the CAS Key Research Program of Frontier Sciences 
and by a Key R\&D Program of Ministry of Science and
Technology under number 2017YFA0402204 and Natural Science Foundation of Shandong Province under the grants ZR2018QA007. BZ is also supported by the Basic Science Research Program through the National Research Foundation of Korea (NRF) funded by the Ministry of Education, Science and Technology (NRF-2019R1A2C2003738), and by the Korea Research Fellowship Program through the NRF funded by the Ministry of Science and ICT (2019H1D3A1A01070937). The~work by M.K. was supported by the Ministry of Science and Higher Education of the Russian Federation as part of the Program for Improving the Competitiveness of the MEPhI (project no. 02.a03.21.0005) and Project "Fundamental problems of cosmic rays and dark matter", No 0723-2020-0040.
\bibliography{refs}

 \end{document}